\documentclass[aps,pra,twocolumn,amsmath,showpacs]{revtex4}
\usepackage{bm}
\usepackage{graphicx}
\begin{document}
\setlength{\arraycolsep}{2pt}
\title{Entanglement condition via su(2) and su(1,1) algebra using Schr{\"o}dinger-Robertson uncertainty relation}
\author{Hyunchul Nha$^*$}
\affiliation{ARC Center of Excellence for Quantum Computer Technology, University of Queensland, Australia\\and\\School of Computational Sciences, Korea Institute for Advanced Study, Korea} 
\date{\today}
%\maketitle
\begin{abstract}
The Schr{\"o}dinger-Robertson inequality generally provides a stronger bound on the product of uncertainties for two noncommuting observables than the Heisenberg uncertainty relation, and as such, it can yield a stricter separability condition in conjunction with partial transposition. In this paper, using the Schr{\"o}dinger-Robertson uncertainty relation, the separability condition previously derived from the su(2) and the su(1,1) algebra is made stricter and refined to a form invariant with respect to local phase shifts. Furthermore, a linear optical scheme is proposed to test this invariant separability condition.\end{abstract}
\maketitle
\email{phylove00@gmail.com}

\narrowtext
\section{Introduction}
When a quantum system is subject to measurements corresponding to two noncommuting observables \{$A$, $B$\}, the product of uncertainties in measurement outcomes, $\langle(\Delta A)^2\rangle\langle(\Delta B)^2\rangle$, has a certain lower bound. The Heisenberg uncertainty relation (HUR)\cite{Heisenberg}, which is most widely used, provides the bound as 
\begin{eqnarray}
\langle(\Delta A)^2\rangle\langle(\Delta B)^2\rangle\ge\frac{1}{4}|\langle[A,B]\rangle|^2.
\label{eqn:HUR}
\end{eqnarray}
On the other hand, the Schr{\"o}dinger-Robertson relation(SRR) \cite{SR1,SR2} in general provides a stronger bound as
\begin{eqnarray}
\langle(\Delta A)^2\rangle\langle(\Delta B)^2\rangle\ge\frac{1}{4}|\langle[A,B]\rangle|^2+\langle\Delta A\Delta B\rangle^2_S,
\label{eqn:SR}
\end{eqnarray}
where the cross correlation $\langle\Delta A\Delta B\rangle_S$ is defined in a symmetric form as
\begin{eqnarray}
\langle\Delta A\Delta B\rangle_S\equiv\frac{1}{2}\langle\Delta A\Delta B+\Delta B\Delta A\rangle.
\end{eqnarray}
The SRR can be derived from the Cauchy-Schwartz inequality, $\langle f|f\rangle\langle g|g\rangle\ge|\langle f|g\rangle|^2$, where $|f\rangle=\Delta A|\Psi\rangle$ and $|g\rangle=\Delta B|\Psi\rangle$ for a generic quantum state $|\Psi\rangle$\cite{Dodonov1}. The HUR describes a special case of the SRR under the condition $\langle\Delta A\Delta B\rangle_S=0$, which is of course not always met.

Recently, one of the important issues in quantum informatics has been to obtain conditions by which one can distinguish entangled states from separable ones. Some of such entanglement criteria derived so far have relied on the bounds set by various forms of uncertainty relations\cite{Hofmann,Guhne,Raymer,Hillery1}, and remarkably for certain cases, in explicit conjunction with partial transposition(PT)\cite{Shchukin1,Agarwal,nha1}. More precisely, separable states can represent a certain physical state even under PT\cite{Peres} and all uncertainty relations must therefore be satisfied by separable states under PT. The uncertainty relations in combination with PT can thereby provide necessary conditions for separability. 

For continuous variables (CVs), earlier works were focused on Gaussian entangled states\cite{Duan,Simon,Mancini}, but considerable attention has also been directed to non-Gaussian entangled states\cite{nha}. Most of all, the separability conditions applicable to non-Gaussian entangled states have recently emerged\cite{Shchukin1,Hillery1,Agarwal,nha1}, and in particular, Refs.~\cite{Agarwal,nha1,Hillery1} employed the su(2) and the su(1,1) algebra to derive such entanglement criteria. Using the HUR along with those two algebras, Nha and Kim have particularly derived the optimal separability condition among a certain class of inequalities\cite{nha1}. This condition has also been proposed to detect multipartite entanglement of photonic $W$ states and shown to be robust against the detector inefficiency\cite{nha2}.

In this paper, it is our aim to refine the separability condition in Refs.~\cite{Agarwal,nha1} by employing the SRR instead of the HUR. By doing this, we obtain a stricter separability condition given by a form invariant with respect to local phase shifts. This invariance is a very adequate attribute as entanglement condition, for entanglement property must be invariant under any local unitary operations. Furthermore, we propose how to experimentally test this invariant condition using linear optics and also discuss the practical connection of the previous condition in \cite{Agarwal,nha1} to the present one.

\section{Separability condition}
First, we briefly introduce how to derive the separability condition via the uncertainty relations in the su(2) and the su(1,1) algebra\cite{nha1}. 
The su(2) algebra deals with the angular momentum operators $J_x,J_y$ and $J_z$, 
which obey the commutation relations $\left[J_i,J_j\right]=i\epsilon_{ijk}J_k$ $(i,j,k=x,y,z)$. 
This algebra can be represented by two bosonic operators $a$ and $b$, 
as 
\begin{eqnarray}
J_x&=&\frac{1}{2}\left(a^\dag b+ab^\dag\right),\nonumber\\ 
J_y&=&\frac{1}{2i}\left(a^\dag b-ab^\dag\right),\nonumber\\ 
J_z&=&\frac{1}{2}\left(a^\dag a-b^\dag b\right).
\label{eqn:su2operators}
\end{eqnarray}
On the other hand, the operators $K_x,K_y$ and $K_z$ in the su(1,1) algebra can be represented by
\begin{eqnarray}
K_x&=&\frac{1}{2}\left(a^\dag b^\dag+ab\right),\nonumber\\ 
K_y&=&\frac{1}{2i}\left(a^\dag b^\dag-ab\right),\nonumber\\ 
K_z&=&\frac{1}{2}\left(a^\dag a+b^\dag b+1\right),
\label{eqn:su11operators}
\end{eqnarray}
which results in the commutation relations, $\left[K_x,K_y\right]=-iK_z,\left[K_y,K_z\right]=iK_x$, and $\left[K_z,K_x\right]=iK_y$, 
different in sign from those of the su(2) algebra.

Specifically, the commutator $\left[K_x,K_y\right]=-iK_z$ in the su(1,1) algebra gives the uncertainty relation via the HUR as
\begin{eqnarray}
\langle(\Delta K_x)^2\rangle\langle(\Delta K_y)^2\rangle\ge\frac{1}{4}|\langle K_z\rangle|^2,
\label{eqn:pi}
\end{eqnarray} 
which must be satisfied by any quantum states.
Most importantly, the inequality~(\ref{eqn:pi}) must be satisfied under PT by every separable state, since it can still describe a certain physical state\cite{Peres}. 
That is, one obtains the separability condition as 
\begin{eqnarray}
\langle(\Delta K_x)^2\rangle_{\rm PT}\langle(\Delta K_y)^2\rangle_{\rm PT}\ge\frac{1}{4}|\langle K_z\rangle|_{\rm PT}^2,
\label{eqn:pi1}
\end{eqnarray} 
where the subscript PT means that the quantum average is calculated after taking partial transposition. 
Using a general relation 
\begin{eqnarray}
\langle a^{\dag m}a^nb^{\dag p}b^q\rangle_{\rho^{\rm PT}}=\langle a^{\dag m}a^nb^{\dag q}b^p\rangle_{\rho} 
\label{eqn:pt}
\end{eqnarray}
between the quantum average for the partially transposed density operator $\rho^{\rm PT}$ and that for the original density operator $\rho$\cite{nha1},  
the inequality~(\ref{eqn:pi1}) can be recast to give the separability condition expressed as 
\begin{eqnarray}
\left[\frac{1}{4}+\langle\left(\Delta J_x\right)^2\rangle\right]\left[\frac{1}{4}+\langle\left(\Delta J_y\right)^2\rangle\right]\ge\frac{1}{16}\left[1+\langle N_+\rangle\right]^2,
\label{eqn:opt}
\end{eqnarray}
where $N_+=a^{\dag}a+b^{\dag}b$ is the total excitation number. 
Note that the inequality~(\ref{eqn:opt}) is the optimal condition derived in \cite{nha1}, where the HUR was employed in a sum form to obtain a class of separability conditions\cite{Eisert}. 
 
\section{stricter separability condition}
In this section, let us now start from the SRR for the commutator $\left[K_x,K_y\right]=-iK_z$, i.e.,
\begin{eqnarray}
\langle(\Delta K_x)^2\rangle\langle(\Delta K_y)^2\rangle\ge\frac{1}{4}|\langle K_z\rangle|^2+\langle\Delta K_x\Delta K_y\rangle^2_S,
\label{eqn:pi2}
\end{eqnarray}
instead of the HUR, then follow the same steps as below Eq.~(\ref{eqn:pi}). Using the relation
\begin{eqnarray}
\langle\Delta K_x\Delta K_y\rangle_{S,\rm PT}=\langle\Delta J_x\Delta J_y\rangle_S
\end{eqnarray} 
via Eq.~(\ref{eqn:pt}), we obtain a separability condition stricter than the one in~(\ref{eqn:opt}) as
\begin{eqnarray}
\left[\frac{1}{4}+\langle\left(\Delta J_x\right)^2\rangle\right]\left[\frac{1}{4}+\langle\left(\Delta J_y\right)^2\rangle\right]&&\nonumber\\
\ge\frac{1}{16}\left[1+\langle N_+\rangle\right]^2&+&\langle\Delta J_x\Delta J_y\rangle^2_S.
\label{eqn:strict}
\end{eqnarray}
Compared with the inequality~(\ref{eqn:opt}), the new inequality~(\ref{eqn:strict}) prodvides a stronger condition for separability as long as the off-diagonal covariance $\langle\Delta J_x\Delta J_y\rangle_S$ is nonzero.
As an example, consider the two-photon entangled states of the type $|\Psi\rangle=\cos\theta|2,0\rangle+i\sin\theta|0,2\rangle$. 
All these states satisfy the inequality~(\ref{eqn:opt}), but violate the stricter one in~(\ref{eqn:strict}), regardless of the parameter $\theta$. 
Therefore, only the inequality~(\ref{eqn:strict}) can detect entanglement for those two-photon states.

We next show that the inequality~(\ref{eqn:strict}) is invariant with respect to local phase shifts. 
Let us consider a $2\times2$ covariance matrix $C$ of which elements are defined as 
\begin{eqnarray}
C_{ij}\equiv\frac{1}{2}\langle\Delta J_i\Delta J_j+\Delta J_j\Delta J_i\rangle,
\end{eqnarray}
where $\{i,j\}=\{x,y\}$. The inequality~(\ref{eqn:strict}) is then expressed as 
\begin{eqnarray}
{\rm Det}\{C\}+\frac{1}{4}{\rm Tr}\{C\}\ge\frac{1}{16}\left(\langle N_+\rangle^2+2\langle N_+\rangle\right),
\label{eqn:mstrict}
\end{eqnarray}
where ${\rm Det}\{\}$ and ${\rm Tr}\{\}$ denote the determinant and the trace of a matrix.
If one takes a local phase shift for mode $b$ as $b'=be^{-i\phi}$, the su(2) operators $J_x$ and $J_y$ are transformed into
\begin{eqnarray}
\begin{pmatrix}&J'_x\\&J'_y
\end{pmatrix}
=\begin{pmatrix}
&\cos\phi&\sin\phi\\&-\sin\phi&\cos\phi
\end{pmatrix}
\begin{pmatrix}&J_x\\&J_y
\end{pmatrix}.
\label{eqn:rotation}
\end{eqnarray}
The determinant and the trace of a matrix are unchanged under rotation, and the total photon number $\langle N_+\rangle$ is also preserved through passive optical elements. 
The inequality~(\ref{eqn:mstrict}) is therefore invariant with respect to local phase shifts. 
This is an attribute very adequate as entanglement condition, for entanglement should be invariant under local unitary operations. 
Note that a phase shift is the only local unitary operation that preserves the total photon number.

\section{Measurement scheme}
We now discuss how the separability condition~(\ref{eqn:strict}) can be tested in experiment. In Ref.\cite{nha1}, a linear optical scheme was proposed to measure the observables $J_x,J_y$ and $\langle N_+\rangle$ for the inequality~(\ref{eqn:opt}), as depicted in Fig.~1. 
The mode $b$ first undergoes a phase shift by $\phi$ and 
the two modes $a$ and $b$ are then injected to a 50:50 beam splitter. 
The modes $c$ and $d$ at the output are given by $c=\frac{1}{\sqrt{2}}(a+be^{-i\phi})$ and 
$d=\frac{1}{\sqrt{2}}(-a+be^{-i\phi})$, respectively. 
One needs to measure the photon number difference at the output, i.e., 
\begin{eqnarray} 
N_{\{-,\phi\}}\equiv c^\dag c-d^\dag d=a^\dag be^{-i\phi}+ab^\dag e^{i\phi},
\end{eqnarray} 
which becomes $2J_x$ ($2J_y$) for $\phi=0$ ($\phi=\frac{\pi}{2}$). (See Eq.~(\ref{eqn:su2operators}).) 
The total photon number $\langle N_+\rangle$ is simply given by the sum, $c^\dag c+d^\dag d$, at the output.  

\begin{figure}
\includegraphics*[width=2.7in,keepaspectratio=true]{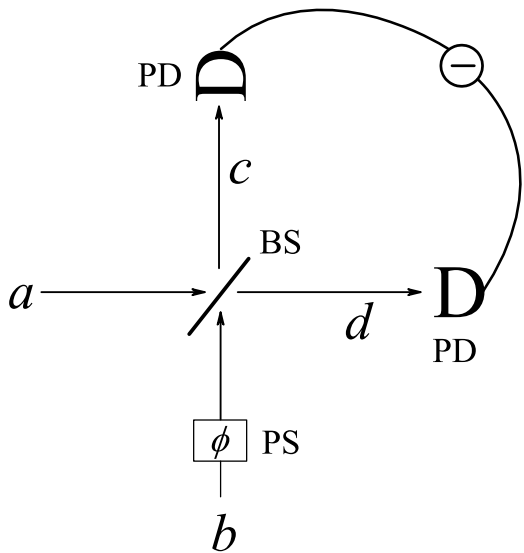}
%\vspace{-2.3in}
\caption{Experimental scheme for measuring the quantities necessary to test the inequality~(\ref{eqn:strict}). 
All the quantum averages in~(\ref{eqn:strict}) can be measured by detecting the photon number difference at the output, $N_{\{-,\phi\}}\equiv c^\dag c-d^\dag d=a^\dag be^{-i\phi}+ab^\dag e^{i\phi}$, with four different phase shifts, $\phi=0,\frac{\pi}{2},\frac{\pi}{4}$, and $-\frac{\pi}{4}$. (See the main text.) BS: 50:50 beam-splitter, PS: phase-shifter, and PD: photo detector.}
\label{fig:fig1}
\end{figure}

In the present inequality~(\ref{eqn:strict}), in addition to $J_x,J_y$ and $\langle N_+\rangle$, one also needs to measure the off-diagonal covariance $\langle\Delta J_x\Delta J_y\rangle_S$. Note that $\langle\Delta J_x\Delta J_y\rangle_S=\frac{1}{2}\langle J_xJ_y+J_yJ_x\rangle-\langle J_x\rangle\langle J_y\rangle$, where
\begin{eqnarray}
J_xJ_y+J_yJ_x&=&\frac{1}{2i}\left(a^{\dag2}b^2-a^2b^{\dag2}\right)\nonumber\\
&=&\frac{1}{4}\left(N_{\{-,\phi=\frac{\pi}{4}\}}^2-N_{\{-,\phi=-\frac{\pi}{4}\}}^2\right).
\label{eqn:mcov}
\end{eqnarray}
Thus, by choosing two different phase shifts $\phi=\frac{\pi}{4}$ and $\phi=-\frac{\pi}{4}$ in Fig.~1, the quantum average $\langle J_xJ_y+J_yJ_x\rangle$ can be measured in two pieces as shown in Eq.~(\ref{eqn:mcov}). In summary, the single experimental setup in Fig.~1 can be used to measure all the quantities necessary to test the inequality~(\ref{eqn:strict}). 

Finally, we discuss how the inequality~(\ref{eqn:opt}) can be regarded as "equivalent" to the stricter inequality~(\ref{eqn:strict}). Using the relation in Eq.~(\ref{eqn:rotation}) implemented by a local phase shift, one has the covariance in the rotated frame as 
\begin{eqnarray}
\langle\Delta J'_x\Delta J'_y\rangle_S=&&\frac{1}{2}\sin2\phi\left[\langle(\Delta J_y)^2\rangle-\langle(\Delta J_x)^2\rangle\right]\nonumber\\&&+\cos2\phi\langle\Delta J_x\Delta J_y\rangle_S.
\end{eqnarray} 
Thus, by choosing the phase shift as 
\begin{eqnarray}
\tan2\phi=\frac{2\langle\Delta J_x\Delta J_y\rangle_S}{\langle(\Delta J_x)^2\rangle-\langle(\Delta J_y)^2\rangle},
\label{eqn:pc}
\end{eqnarray} 
the covariance in the rotated frame can be made vanish. In this situation, the inequality~(\ref{eqn:strict}) is reduced to the inequality~(\ref{eqn:opt}). In other words, as long as one is allowed to perform a local phase shift, which does not alter the entanglement property at all, the two inequalities can be interpreted as equivalently useful. However, this relies on the capability of measuring all the covariances and of performing a phase shift very accurately required by Eq.~(\ref{eqn:pc}). It is then of no practical advantage to adhere to the inequality~(\ref{eqn:opt}): One can simply test the inequality~(\ref{eqn:strict}) if one is able to measure the off-diagonal covariance $\langle\Delta J_x\Delta J_y\rangle_S$ in addition. 
 
\section{Summary}
In this paper, we have derived a stricter separability condition via the su(2) and the su(1,1) algebra using the Schr{\"o}dinger-Robertson inequality instead of the Heisenberg uncertainty relation. It has been shown that this refined condition is expressed in a form invariant with respect to local phase shifts. A linear optical setup has been proposed to test the invariant separability condition and the practical connection of the previously obtained condition to the present one was also discussed. 

{\it Note added in proof.} Recently, the author has learned that a similar linear optical method was proposed to measure the same quantities as the ones in this paper, but in a different context\cite{Campos}. 
\section{Acknowledgment}
This work was supported by the University of Queensland.

*email:phylove00@gmail.com

\end{document}